\documentclass[aps,prd,10pt,showpacs,amsmath,showkeys,twocolumn,floatfix,amssymb, preprintnumbers, nofootinbib, superscriptaddress]{revtex4}
\usepackage{graphicx}
\usepackage{comment}
\usepackage[usenames]{color}
\usepackage{bm}
\usepackage{ifpdf}
\usepackage{floatrow}
\usepackage{makecell}

\usepackage[normalem]{ulem}
\usepackage[dvipsnames]{xcolor}
\usepackage[utf8]{inputenc}
\usepackage{hyperref}
\hypersetup{
  pdfnewwindow=true,      
  colorlinks=true,        
  linkcolor=PineGreen,    
  citecolor=PineGreen,    
  filecolor=PineGreen,    
  urlcolor=PineGreen      
}
\newcommand{\be}{\begin{eqnarray}}
\newcommand{\ee}{\end{eqnarray}}

\begin{document}

\title{Nuclear reactions in artificial traps}

\author{Peng~Guo}
\email{pguo@csub.edu}

\affiliation{College of Physics, Sichuan University, Chengdu, Sichuan 610065, China}
\affiliation{Department of Physics and Engineering,  California State University, Bakersfield, CA 93311, USA}
\affiliation{Kavli Institute for Theoretical Physics, University of California, Santa Barbara, CA 93106, USA}

\author{Bingwei~Long}
\email{bingwei@scu.edu.cn}
\affiliation{College of Physics,  Sichuan University, Chengdu, Sichuan 610065, China}

\date{\today}

\begin{abstract}
Coupled-channel two-particle systems bound by a harmonic trap are discussed in the present paper. We derive the formula that relates the energy levels of such trapped systems to phase shifts and inelasticity of coupled-channel reactions. The formula makes it possible to extract amplitudes of inelastic nuclear reactions from ab initio calculations of discrete levels of many-nucleon systems in a harmonic trap.
\end{abstract}

\maketitle

\section{Introduction}\label{sec:intro}

Studies of nuclear structure have entered a stage where bound states of many nucleons can be solved for with microscopic nuclear forces and with high precision~\cite{Navratil:2007we, Epelbaum:2008ga, Ekstrom:2013kea, Barrett:2013nh, Lynn:2014zia, Ekstrom:2015rta, Piarulli:2017dwd, Hammer:2019poc}. For some of those many-body solvers, the computation typically involves expansion of wave functions on a chosen basis, e.g., the harmonic-oscillator (HO) basis. This poses a challenge for immediate extension of basis-expansion methods to continuum-state problems like resonances and reactions. The need for synergy of bound-state and continuum computational methods has been recognized by the nuclear-theory community~\cite{Navratil:2016ycn, Yang:2016brl, Johnson:2019sps, Mazur:2020hdc, Ma:2020roi}.

One of these ab initio methods for elastic scattering of two clusters, normally stable nuclei, utilizes the energy spectrum of all the constituent nucleons, trapped by an artificial field, to generate directly the phase shifts at certain discrete energies decided by the spectrum of the trapped~\cite{Rotureau:2010uz, Rotureau:2011vf, Luu:2010hw, Zhang:2020rhz}. This method is mostly enabled by a set of so-called quantization conditions (QCs) that map energy levels to the phase shifts at momenta related to those energy eigenvalues~\cite{Busch98, Zhang:2019cai}. The harmonic trap is an attractive choice for the artificial field because the HO wave functions are analytically known, so such mapping formulas can be presented in closed forms to a great extent. Existing vast resource of software for the HO basis is also an important incentive~\cite{Navratil:2000gs}. After the pioneering work in Ref.~\cite{Busch98} to connect the spectrum of the trapped two-body systems to their elastic scattering amplitude, referred to as the BERW formula in the present paper, more theoretical works to extend the formula and applications to nucleon-nucleon and nucleon-cluster scattering followed~\cite{Stetcu:2007ms, Stetcu:2010xq, Rotureau:2010uz, Rotureau:2011vf,  Luu:2010hw, Yang:2016brl, Zhang:2019cai, Zhang:2020rhz}.  Here we turn our attention to coupled-channel problems where at least one of the two participating particles changes its species after the reaction. This is relevant to nuclear reactions where the nuclei fuse and then break up to different nucleus than before.

Relating discrete energy spectrum of trapped composite particles to their scattering amplitudes has been a topic of great interests in other fields of strong-interaction physics. In these applications, the form of traps may differ, such as periodic cubic boxes in lattice quantum chromodynamics (QCD) \cite{Luscher:1990ux} and spherical hard wall in some of the lattice implementations of chiral effective field theory  \cite{Bovermann:2019jbt,Elhatisari:2016hby,Rokash:2015hra}. Because the underlying theory to compute the energy spectrum is model-independent and the computation often consumes much computing resources, it is highly desired that the said connection is constructed without uncontrolled modeling of interactions between the composite particles.
This can be achieved when two scales are well separated; one is the wideness of the trap $b$ and the other the interaction range between the composite particles $R$, $b \gg R$.
The separation makes it possible to locate a domain inside the trap where the composite particles do not interact appreciably while the confining force is sufficiently weak so that the trapped particles' wave function resembles that of the scattering state in infinite volume. One then matches the asymptotic form of the scattering wave function, which has undetermined scattering parameters, say, phase shifts, to the trapped wave function.
Doing so pins down the phase shift at the particular energy associated with the trapped wave function. This yields the QC, a compact relation between the discrete energy eigenvalues and the scattering parameters, which often takes the following schematic form:
\begin{equation}
   \det \left [  \cot \delta  (E) -  g (E ; b) \right ]_{E=\mathcal{E}_k}=0 \, ,
    \label{eqn:generalQC}
\end{equation}
where $\mathcal{E}_k$ is the $k$-th the energy eigenvalue and
$\delta (E)$ refers to the diagonal matrix with phase shifts as the elements. The analytic matrix function
$g(E; b)$  depends only on the configuration of the trap, such as the functional form of the confining force and the wideness $b$, but not on detail of the inter-particle interactions.

Matching wave functions was the fundamental methodology used by L\"uscher to derive his seminal QC formula for periodic cubic boxes~\cite{Luscher:1990ux}.
Once we are convinced that the QC is irrespective of the interactions, manipulation of wave functions, however, does not have to be the ubiquitous technique of deriving the QC. Actively researched in the field of lattice QCD, other methods have been developed to tackle the problem (see, e.g., Refs.~ \cite{Rummukainen:1995vs,Christ:2005gi,Bernard:2008ax,He:2005ey,Lage:2009zv,Doring:2011vk,Guo:2012hv,Guo:2013vsa,Kreuzer:2008bi,Polejaeva:2012ut,Hansen:2014eka,Mai:2017bge,Mai:2018djl,Doring:2018xxx,Guo:2016fgl,Guo:2017ism,Guo:2017crd,Guo:2018xbv,Guo:2018ibd,Guo:2019hih,Guo:2019ogp,Guo:2020wbl,Guo:2020kph,Guo:2020iep,Guo:2020ikh,Guo:2020spn,Guo:2021lhz}). We will follow the variational approach spelled out in Refs.~\cite{Guo:2020iep,Guo:2020spn,Guo:2021lhz}. In particular, contact interactions are used in describing dynamics in both harmonic trap and infinite volume,   and the   QC is derived by eliminating the contact coupling.

The paper is organized as follows.  To establish our notation, derivation of the BERW formula  is  discussed in Sec.~\ref{sec:berw}. The extension to coupled-channel reactions is presented in Sec.~\ref{sec:coupleberw}, followed by discussions and summary in Sec.~\ref{summary}.

\section{BERW formula\label{sec:berw}}
 
Using the approach in Refs.~\cite{Guo:2020iep,Guo:2020spn,Guo:2021lhz}, we re-derive the BERW formula for elastic scattering of two spinless particles in this section. We consider cases where the kinematics is such that the reaction does not takes places near threshold. So the long-range Coulomb repulsion, if any, can be neglected and one is left with only the strong interaction, which is finite-ranged. 

\subsection{Scattering dynamics}
In infinite volume, the Lippmann-Schwinger (LS) equation for two-particle scattering   has the following operator form,
\begin{equation}
\hat{T}_\infty(E) = -\hat{V} + \hat{V} \hat{G}_\infty(E) \hat{T}_\infty(E) \, , \label{eqn:LSE}
\end{equation}
where $\hat{T}_\infty$ is the $T$-matrix, $E$ the energy, and $\hat{V}$ the short-range interaction between the clusters. The infinite-volume Green's function is given in the operator form by
\begin{equation}
\hat{G}_\infty(E) = \frac{1}{E - \hat{H}_0  } \, ,
\end{equation}
with $\hat{H}_0$ defined in terms of the reduced mass $\mu$ and the Laplacian:
\begin{equation}
    \hat{H}_0 = -\frac{\nabla^2}{2\mu} \, .
\end{equation}
The solution to the LS equation can be formally written as
\begin{equation}
    \hat{T}_\infty(E) = - \left[\hat{V}^{-1} -  \hat{G}_\infty(E) \right]^{-1} \, . \label{Tinf}
\end{equation}
 
The on-shell elements of partial-wave projected $T$-matrix are usually parameterized in the center-of-mass (CM) frame by the phase shift $\delta_L$ as
\begin{equation}
T^{(\infty)}_{L}(q, q) = \frac{(4\pi)^2}{2 \mu q} \frac{1}{  \cot \delta_L(q) - i } \, ,
\end{equation}
where $q=\sqrt{2\mu E}$ is relative momentum of particles.

\subsection{Two particles trapped by a HO potential}
One may still define  the $T$-matrix  for the trapped two-particle system for any values of $E$, even though there are no scattering states:
\begin{equation}
    \hat{T}_{\omega}(E) = -\hat{V} +  \hat{V}  \hat{G}_{\omega}(E) \hat{T}_{\omega}(E) \, , \label{Tho}
\end{equation}
where  the Green's function of the HO potential with angular frequency $\omega$ is given by
\begin{equation}
     \hat{G}_{\omega}(E) = \frac{1}{E - \hat{H}_\omega } = \sum_n \frac{|n \rangle \langle n| }{E - \omega (n+ \frac{3}{2}) } \, .
\end{equation}
Here $\hat{H}_\omega$ is the HO Hamiltonian,
\begin{equation}
   \hat{H}_\omega  =-\frac{\nabla^2}{2\mu}    +   \frac{1}{2 }  \mu \omega^2 \mathbf{ r}^2 \, ,
\end{equation}
and $ |n \rangle $ stands for eigenstates of  $ \hat{H}_\omega$ operator: 
\begin{equation}
 \hat{H}_\omega |n \rangle  =  \omega(n+ \frac{3}{2})|n \rangle  ,  \ \ \ \ n=0,1,2\cdots.
\end{equation}
The wideness of the harmonic trap can be characterized by the so-called oscillator length:
\begin{equation}
  b \equiv \left(\frac{1}{2} \mu \omega\right)^{-\frac{1}{2}} \, .
\end{equation}

The bound state sustained by the harmonic trap can be determined by poles of $T_\omega(E)$, and
\begin{equation}
\hat{T}_\omega(E) =  - \left[\hat{V}^{-1} -  \hat{G}_\omega (E) \right]^{-1} \, .
\end{equation}
Therefore, any eigenvalue of the full Hamiltonian $(\hat{H}_\omega + \hat{V}  )$ $E = \mathcal{E}_k$ satisfies
\begin{equation}
    \det \left[ \hat{V}^{-1} -  \hat{G}_\omega( E) \right]_{E = \mathcal{E}_k} = 0 \label{eqn:detVGw} \, .
\end{equation}

\subsection{Contact potentials and the BERW formula}
Combining Eqs.~\eqref{Tinf} and \eqref{eqn:detVGw} by eliminating $\hat{V}$,
one finds symbolically
\begin{equation}
\det \left [ \hat{T}_\infty^{-1}(E) + \hat{G}_\omega( E) - \hat{G}_\infty (E) \right]_{E = \mathcal{E}_k} = 0. \label{Tsingle}
\end{equation}
Although the above equation is formally correct, it is difficult to use in practical calculations because inverting $T_\infty(E)$ directly is highly non-trivial. 

As long as $R/b$ is small, as we argued, a model-independent QC must exist. So one can use whatever form of two-body interaction to deduce the QC. We choose a sum of contact interactions, with one term responsible for each partial wave, to describe the cluster-cluster interaction:
\begin{equation}
    V(\mathbf{q}', \mathbf{q}\,) = \sum_L  \frac{2L+1}{4\pi}  (q' q)^L V_L P_L(\hat{\mathbf{q}}'\cdot\hat{\mathbf{q}}) \, , \label{eqn:contactVLO}
\end{equation}
where $P_L(x)$ is the Legendre polynomial and $\mathbf{q}'$ ($\mathbf{q}$) is the outgoing (incoming) momentum.

In the CM frame for infinite volume, the separable form of the potential in Eq.(\ref{eqn:contactVLO}) allows for a closed-form partial-wave solution to Eq.(\ref{Tinf}) (see also Refs.~\cite{Guo:2020spn,Guo:2021lhz}):
\begin{align}
& - \frac{2\mu q^{2L+1}}{(4\pi)^2} \left [ \cot \delta_L(q) - i  \right ]  \nonumber \\
& = \frac{1}{V_L} -   \frac{2^{2L+1} \Gamma^2(L+\frac{3}{2})}{(2\pi)^3} \frac{G^{(\infty)}_L (r,r'; q)}{(r r')^L} |_{r,r' \rightarrow 0}  \, , \label{eqn:invTsingle}
\end{align}
where  
$$ G^{( \infty)}_L (r, r'; q   )  =    - i 2\mu q   j_L (q  r_<) h_L^{(+)} ( q  r_>) $$
is the free-particle Green's function in the $L$-th partial wave.

In the harmonic trap, the separable form also facilitates a closed-form solution to Eq.(\ref{eqn:detVGw}):
\begin{equation}
\frac{1}{V_L } =      \frac{2^{2L+1} \Gamma^2(L+\frac{3}{2})}{(2\pi)^3} \frac{G^{(\omega)}_L (r,r'; \epsilon^{(\omega)})}{(r r')^L} |_{r,r' \rightarrow 0}  \, ,   \label{eqn:homoVsingle}
\end{equation}
where $G^{(  \omega)}_L$ is the partial-wave HO Green's function and is given in Ref.~\cite{Blinder83} by
 \begin{align}
 & G^{(  \omega)}_L (r, r'; \epsilon^{(\omega)})  = - \frac{1}{\omega (r r')^{ \frac{3}{2}}} \frac{ \Gamma (\frac{L}{2} + \frac{3}{4} - \frac{ \epsilon^{(\omega)}}{2 \omega}) }{\Gamma(L+\frac{3}{2})}  \nonumber \\
& \times  \mathcal{M}_{\frac{\epsilon^{(\omega)}}{2\omega}, \frac{L}{2} + \frac{1}{4} }( \mu \omega r^2_{<}) \mathcal{W}_{\frac{\epsilon^{(\omega)}}{2 \omega},\frac{L}{2} + \frac{1}{4} }  (\mu \omega r^2_{>}) \, . \nonumber
 \end{align}
$ \mathcal{M}$ and $\mathcal{W}$ are the Whittaker functions as defined in  Ref.~\cite{whittaker_watson_1996}. $\epsilon^{(\omega)} $ represents the energy shift relative to the HO levels and is related to relative momentum $q$ by $$\frac{q^2}{2\mu} =  \epsilon^{(\omega)} + \omega (n+ \frac{3}{2})=E\, .$$
where $\omega (n+ \frac{3}{2})$ is the CM energy in the harmonic trap, and the CM frame in infinite volume is assumed.
 
By eliminating $V_L$ from Eqs.~\eqref{eqn:invTsingle} and \eqref{eqn:homoVsingle}, and using asymptotic forms of partial-wave Green's functions $G^{(  \omega)}_L$ and  $G^{( \infty)}_L$, one obtains the BERW formula:
\begin{equation}
  \cot \delta_L (q ) = g_L(q) ,
\end{equation}
where
\begin{equation}
 g_L(q) = (-1)^{L+1} \left ( \frac{4 \mu \omega}{q^2} \right )^{L+ \frac{1}{2}}         \frac{ \Gamma ( \frac{3}{4} + \frac{L}{2} - \frac{ \epsilon^{(\omega)}}{2 \omega}) }{\Gamma ( \frac{1}{4}-\frac{L}{2} - \frac{\epsilon^{(\omega)}}{2\omega} )} . \label{gLsingle}
\end{equation}

\section{Two-channel reactions \label{sec:coupleberw}}

For nuclear reactions where two nuclei fuse, break up, and rearrange, the BERW formula can not be immediately applied. Extension of the derivation in Sec.~\ref{sec:berw} is needed to extract inelastic scattering amplitudes. We focus on reactions that involve two channels and assume that each channel is made up of two distinguishable particles $A + B$ and $C + D$. Labelling two channels by $1$ and $2$, we can write, for instance, reaction $1 \to 2$ as
\begin{equation}
  A + B \to C + D \, .
\end{equation}
The kinetic energies in both channels are related by 
\begin{equation}
    \frac{\mathbf{q}_1^2}{2\mu_1} + \frac{\mathbf{ P}^2}{2M_1}= \frac{\mathbf{q}_2^2}{2\mu_2} + \frac{\mathbf{ P}^2}{2M_2}+\Delta =E \, ,
\end{equation}
where  $\mathbf{P}$ is the total momentum, $\mathbf{q}_{1, 2}$ relative momenta, $M_{1, 2}$ the total mass of each channel, $$\Delta \equiv M_2 - M_1,$$ and $\mu_{1,2}$ the reduced mass. For simplicity, we will assume the CM frame in what follows: $\mathbf{P} = 0$.

\subsection{Coupled-channel energy levels in a harmonic trap}

We start with the coupled Schr\"odinger equations for two interacting spinless particles bound by the HO potential:
\begin{equation}
\left [ \hat{H}_{\omega}  +  V (\mathbf{ r} ) \right ]
\begin{bmatrix}
\psi^{(1, \omega)}  ( \mathbf{ r}  )  \\
\psi^{(2, \omega)}  ( \mathbf{ r}  )
\end{bmatrix}
 =\begin{bmatrix}
   \epsilon^{(\omega)}_1 & 0  \\
 0&  \epsilon^{(\omega)}_2 
\end{bmatrix}    \begin{bmatrix}
\psi^{(1, \omega)} ( \mathbf{ r}  )  \\
\psi^{(2, \omega)}  ( \mathbf{ r}  )
\end{bmatrix} , \label{schpsicoup}
\end{equation}
where 
$\hat{H}_{\omega}$ is given by
   \begin{equation}
   \hat{H}_{\omega}  = \begin{bmatrix}
   - \frac{\nabla_{\mathbf{ r} }^2}{2 \mu_1 }  + \frac{1}{2} \mu_1 \omega^2 \mathbf{ r}^2 & 0  \\
 0&  - \frac{\nabla_{\mathbf{ r} }^2}{2 \mu_2 }  + \frac{1}{2} \mu_2 \omega^2 \mathbf{ r}^2 
\end{bmatrix} .
   \end{equation}
Here $\mathbf{ r}$ is the relative coordinate and the potential matrix $V (\mathbf{ r} )$ is short-ranged and describes inter-cluster interactions and transition between the channels:
\begin{equation}
   V (\mathbf{ r} ) =
    \begin{bmatrix}
   V^{(1,1)} (\mathbf{ r} )& V^{(1, 2)} (\mathbf{ r}) \\
V^{(2,1)} (\mathbf{ r} )& V^{(2, 2)} (\mathbf{ r})
\end{bmatrix}.
   \end{equation}
$ \epsilon^{(\omega)}_1   $ and $ \epsilon^{(\omega)}_2  $ are 
energy shifts in their respective channel and  are   related to the full energy $E$ by
$$\epsilon^{(\omega)}_1  + \omega (n_1 + \frac{3}{2})   =  \epsilon^{(\omega)}_2 + \omega (n_2 + \frac{3}{2}) +\Delta = E\, .$$  

The integral representation of Eq.(\ref{schpsicoup}) is  the coupled LS equation:
   \begin{equation}
  \begin{bmatrix}
\psi^{(1, \omega)} ( \mathbf{ r}  )  \\
\psi^{(2, \omega)}  ( \mathbf{ r}  )
\end{bmatrix}  =
    \int d \mathbf{ r}'  
   G^{( \omega)} (\mathbf{ r}, \mathbf{ r}' ;  E)  
V  (\mathbf{ r}' )
\begin{bmatrix}
\psi^{(1, \omega)}  ( \mathbf{ r}'  )  \\
\psi^{(2, \omega)}  ( \mathbf{ r}'  )
\end{bmatrix},\label{eqn:CoupledHomo_rspace}
\end{equation}
where $ G^{( \omega)}$ denotes the coupled-channel version of the HO Green's function:
   \begin{align}
  G^{( \omega)} (\mathbf{ r}, \mathbf{ r}' ;  E)  =
      \begin{bmatrix}
  G^{(1, \omega)} (\mathbf{ r}, \mathbf{ r}'; \epsilon^{(\omega)}_1 ) & 0   \\
 0 &   G^{(2, \omega)} (\mathbf{ r}, \mathbf{ r}'; \epsilon^{(\omega)}_2 )
\end{bmatrix}  ,
\end{align}
and    $ G^{(\alpha, \omega)}$    satisfies
\begin{equation}
\left [ \epsilon^{(\omega)}_\alpha + \frac{\nabla_{\mathbf{ r} }^2}{2 \mu_\alpha }  - \frac{1}{2} \mu_\alpha \omega^2 \mathbf{ r}^2 \right ] G^{(\alpha,\omega)} (\mathbf{ r}, \mathbf{ r}'; \epsilon^{(\omega)}_\alpha)
   =  \delta(\mathbf{ r} -\mathbf{ r}').
\end{equation}
The partial-wave projection of the HO Green's function is similar to the uncoupled-channel case:
\begin{equation}
  G^{(\alpha,\omega)} (\mathbf{ r}, \mathbf{ r}'; \epsilon^{( \omega)}_\alpha)  = \sum_{L}\frac{2L+1}{4\pi}   G^{(\alpha,\omega)}_L (r, r'; \epsilon^{( \omega)}_\alpha) P_{L} (\mathbf{ \hat{r}} \cdot \mathbf{ \hat{r}}'),   
  \end{equation}
  where
  \begin{align}
 &G^{(\alpha, \omega)}_L (r, r'; \epsilon^{( \omega)}_\alpha)  = - \frac{1}{\omega (r r')^{ \frac{3}{2}}} \frac{ \Gamma (\frac{L}{2} + \frac{3}{4} - \frac{ \epsilon^{( \omega)}_\alpha}{2 \omega}) }{\Gamma(L+\frac{3}{2})}  \nonumber \\
& \times  \mathcal{M}_{\frac{\epsilon^{( \omega)}_\alpha}{2\omega}, \frac{L}{2} + \frac{1}{4} }( \mu_\alpha \omega r^2_{<}) \mathcal{W}_{\frac{\epsilon^{( \omega)}_\alpha}{2 \omega},\frac{L}{2} + \frac{1}{4} }  (\mu_\alpha \omega r^2_{>}) .
 \end{align}

We are going to take one more step to write the integral equation~\eqref{eqn:CoupledHomo_rspace} in momentum space:
 \begin{equation}
  T^{(  \omega)} (  \mathbf{ q}'  )  =  \int \frac{ d \mathbf{ q} }{(2\pi)^3}      \frac{d \mathbf{ k}}{(2\pi)^3}    \widetilde{V} ( \mathbf{ q}' - \mathbf{ k} )       \widetilde{G}^{(  \omega)} (\mathbf{ k}, \mathbf{ q};  E) T^{( \omega)} ( \mathbf{ q}  ) , \label{LScouple}
  \end{equation}
where      $T^{( \omega)} (  \mathbf{ q}'  ) $   matrix is defined by
 \begin{equation}
 T^{( \omega)} ( \mathbf{ q}'  )   = - \int d \mathbf{ r}
 e^{- i \mathbf{ q}' \cdot \mathbf{ r}}
 V (\mathbf{ r} ) \begin{bmatrix}
\psi^{(1, \omega)} ( \mathbf{ r}  )  \\
\psi^{(2, \omega)} ( \mathbf{ r}  )
\end{bmatrix}.
\end{equation}
$ \widetilde{V}$ and $  \widetilde{G}^{(  \omega)} $ are the Fourier transform of $V$ and $G^{(  \omega)} $,  respectively. 
 The partial wave expansion of Eq.(\ref{LScouple}) yields
  \begin{equation}
  T^{( \omega)}_{ L} ( q'  )   =  \int_0^\infty \frac{  q^2 d  q }{(2\pi)^3}   \frac{  k^2 d  k }{(2\pi)^3}        \widetilde{V}_L (  q', k)    \widetilde{G}^{( \omega)}_L ( k , q;  E )   T^{( \omega)}_{ L } ( q ). \label{LScouplePW}
\end{equation}
 
With the assumption of separable potential, 
\begin{equation}
\widetilde{V}_L ( q'  ,q ) = (q' q)^L V_L \, , \label{eqn:CpldVL}
\end{equation}
where $V_L$ is a constant $2\times 2$ matrix, it follows that $T^{(  \omega)}_{L}( q' ) $ matrix must be of the form
$$ T^{(  \omega)}_{L}( q ) = q^L  t^{(\omega)}_{ L} \, ,$$
where $t^{(\omega)}_L$ matrix does not depend on $q$.
Therefore, the quantization condition is given by
\begin{equation}
\det \left [ V_L^{-1} -  \int_0^\infty \frac{  q^2 d  q }{(2\pi)^3}  \frac{  k^2 d  k }{(2\pi)^3}     (k q)^L
 \widetilde{G}^{( \omega)}_L ( k , q;  E )   \right ]=0
. \label{eqn:Vcouple_apdx}
\end{equation}

Using the limiting form of the spherical Bessel functions
\begin{equation}
q^L =\frac{2^{L+1} \Gamma(L+\frac{3}{2})}{\sqrt{\pi}   } \frac{j_L (q r)}{r^L} |_{ r\rightarrow 0}  \, , \label{besselJasym}
\end{equation}
one can show that
\begin{align}
 & \int_0^\infty \frac{  q^2 d  q }{(2\pi)^3} \frac{  k^2 d  k }{(2\pi)^3}    (kq)^L \widetilde{G}^{(\omega)}_L ( k, q; E )  \nonumber \\
 &  =        \frac{2^{2L+1} \Gamma^2(L+\frac{3}{2})}{(2\pi)^3} \frac{G^{(\omega)}_L (r,r'; E)}{(r r')^L} |_{r,r' \rightarrow 0}\, .  
 \end{align}
Therefore, we can rewrite Eq.(\ref{eqn:Vcouple_apdx}) as
 \begin{equation}
\det \left [ V_L^{-1} -        \frac{2^{2L+1} \Gamma^2(L+\frac{3}{2})}{(2\pi)^3  } \frac{G^{(\omega)}_L (r,r'; E)}{ (r r')^L} |_{r,r' \rightarrow 0}   \right ]=0
, \label{eqn:Vcouple_coord}
\end{equation}
which is just the generalization of Eq.(\ref{eqn:homoVsingle}) in the case of coupled-channel reactions.

\subsection{
Coupled-channel reaction amplitude}
 In infinite volume,  the coupled-channel reaction is described by the two-channel LS equation:
 \begin{align}
&  T^{( \infty)}( \mathbf{ q}' , \mathbf{ q} ) =  - \widetilde{V} ( \mathbf{ q}' - \mathbf{ q}  )  \nonumber \\
&+  \int \frac{ d \mathbf{ k} }{(2\pi)^3}     \widetilde{V} ( \mathbf{ q}' - \mathbf{ k} )   \widetilde{G}^{( \infty)} (  \mathbf{ k};  E )    T^{( \infty)}( \mathbf{ k} , \mathbf{ q} )\, , \label{TLSeq}
 \end{align}
 where  $T^{( \infty)}$ is a $2 \times 2$ matrix, 
  \begin{equation}
  T^{( \infty)}   =
    \begin{bmatrix}
   T^{((1,1),\infty)}   & T^{((1, 2),\infty)}     \\
T^{((2,1),\infty)}      & T^{((2, 2),\infty)}    
\end{bmatrix}
\, ,
   \end{equation}
and the free Green's function is also in  matrix form,
 \begin{equation}
 \widetilde{G}^{(\infty)} (\mathbf{ k}; E  )  =
    \begin{bmatrix}
   \frac{1}{\frac{q_1^2}{2\mu_1} -  \frac{\mathbf{ k}^2}{2\mu_1} }& 0   \\
 0 &   \frac{1}{\frac{q_2^2}{2\mu_2} -  \frac{\mathbf{ k}^2}{2\mu_2} }
\end{bmatrix} ,
\end{equation}
with
  $$  \frac{q_1^2}{2\mu_1}  = \frac{q_2^2}{2\mu_2}+\Delta  =E  \, .$$
The partial-wave expansion of  Eq.(\ref{TLSeq}) produces
 \begin{align}
  & T_L^{(\infty)} (q' , q) =  - \widetilde{V}_L ( q'  ,q )  \nonumber \\
  &+  \int_0^\infty \frac{ k^2 d k }{(2\pi)^3}      \widetilde{V}_L ( q' ,  k  )   \widetilde{G}^{(\infty)} ( k ;  E)    T_L^{( \infty)}( k ,q ). \label{LScouplePWinf}
 \end{align}
The separable form of the potential~\eqref{eqn:CpldVL} enables semi-closed form solutions to the above integration equation:
\begin{equation}
  T_L^{( \infty)} (q' , q)  =  (q' q)^L   t_L   (E),
\end{equation}
where  
\begin{equation}
t_L^{-1} (E) = - V_L^{-1} + \int_0^\infty \frac{ k^2 d k }{(2\pi)^3}  k^{2L} \widetilde{G}^{(\infty)} ( k ; E  )   .\label{Vcoupleinf}
\end{equation}
When it is on-shell, the matrix $t_L$ is  usually parameterized by phase shifts, $\delta_L^{(1,2)}$, and inelasticity $\eta_L $,
 \begin{equation}
    \frac{  t_L (E) }{ (4\pi)^2 } = \begin{bmatrix}  \  \frac{ \frac{1}{2i} \left( \eta_L  e^{2 i  \delta_L^{(1)}   }-1 \right)}{ 2\mu_1     q_1^{2L+1}   } &      \frac{ \frac{ \sqrt{1- \eta^2_L}   }{2}e^{i\left (\delta_L^{(1)}   + \delta_L^{(2)}    \right )}}{ 2\mu_1 q_1 (  q_1  q_2)^{L}   }  \\
    \frac{  \frac{ \sqrt{1- \eta^2_L}  }{2} e^{i\left (\delta_L^{(1)}   + \delta_L^{(2)}    \right )}}{ 2\mu_2 q_2 ( q_1 q_2)^{L}   }  &       \frac{ \frac{1}{2i } \left(\eta_L  e^{2 i  \delta_L^{(2)}   }-1 \right) }{ 2\mu_2 q_2^{2L+1}  }
 \end{bmatrix}  . \label{tmat}
\end{equation}

Using the following identity,
 \begin{align}
 & \int_0^\infty \frac{  k^2 d  k }{(2\pi)^3}  k^{2 L}  \widetilde{G}^{(\infty)} ( k ; E )      \nonumber \\
 & =       \frac{2^{2L+1} \Gamma^2(L+\frac{3}{2})}{(2\pi)^3} \frac{G^{(\infty)}_L (r,r'; E)}{(r r')^L} |_{r,r' \rightarrow 0} ,
\end{align}
where     
\begin{align}
 & G^{(\infty)}_L (r, r'; E  ) = 
    \begin{bmatrix}
  G^{(1,\infty)}_L (r, r'; q_1  ) & 0   \\
 0 &   G^{(2,\infty)}_L (r, r'; q_2 )
\end{bmatrix} , \nonumber \\
 & G^{(\alpha,\infty)}_L (r, r'; q_\alpha   )  =    - i 2\mu_\alpha q_\alpha   j_L (q_\alpha  r_<) h_L^{(+)} ( q_\alpha  r_>) \, ,
\end{align}
we find the coupled-channel version of Eq.(\ref{eqn:invTsingle}):
\begin{equation}
t_L^{-1} (E) = - V_L^{-1} +      \frac{2^{2L+1} \Gamma^2(L+\frac{3}{2})}{(2\pi)^3} \frac{G^{(\infty)}_L (r,r'; E)}{(r r')^L} |_{r,r' \rightarrow 0}  .\label{Vcoupleinf_coord}
\end{equation}

\subsection{BERW formula for coupled-channel reactions}

Using Eq.(\ref{Vcoupleinf_coord}) in Eq.(\ref{eqn:Vcouple_coord}) and , one finds
\begin{align}
 \det & \bigg [    \frac{2^{2L+2} \Gamma^2(L+\frac{3}{2})}{\pi}
      \frac{G^{( \omega)}_L (r,r'; E) - G^{( \infty)}_L (r,r'; E) }{(r r')^L}|_{r,r' \rightarrow 0}  \nonumber \\
      &+  (4\pi)^2  t_L^{-1}  (E)  \bigg ]=0 \, . \label{qccoupgreen}
\end{align}
The cancellation of ultraviolet divergences manifested by the Green's functions is made clear if we use 
the limiting forms of both Green's functions:
\begin{align}
&      \frac{2^{2L+2}\Gamma(L + \frac{3}{2})^2 }{\pi }     \frac{G^{(\alpha, \omega)}_{ L} (r,r'; \epsilon^{( \omega)}_\alpha )  }{(r r')^L} |_{r,r' \rightarrow 0}  \nonumber \\
& = -   \frac{ (\mu_\alpha \omega)^{L+ \frac{3}{2}}}{  \omega  }       2^{2L+2}  (-1)^{L+1}   \frac{ \Gamma (\frac{L}{2} + \frac{3}{4} - \frac{ \epsilon^{( \omega)}_\alpha}{2 \omega}) }{\Gamma ( \frac{1}{4}-\frac{L}{2} - \frac{\epsilon^{( \omega)}_\alpha}{2\omega} )} \nonumber \\
&  -         \frac{2^{2L+1} \Gamma(L + \frac{1}{2}) \Gamma(L + \frac{3}{2})}{\pi }   \frac{ 2\mu_\alpha   }{ r^{2L + 1}}    |_{r\rightarrow 0},  \label{asymp1}
 \end{align}
 and
\begin{align}
  &    \frac{2^{2L+2} \Gamma(L + \frac{3}{2})^2}{\pi}    \frac{  G_L^{(\alpha,\infty)} (r,r'; q_\alpha )  }{(r r')^L} |_{r,r' \rightarrow 0}   \nonumber \\
     &  = - i 2 \mu_\alpha q_{\alpha}^{2L+1}       -      \frac{2^{2L+1}  \Gamma(L+ \frac{1}{2}) \Gamma(L + \frac{3}{2})}{\pi}  \frac{2 \mu_\alpha}{r^{2L+1}}  |_{r\rightarrow 0} \, . \label{asymp2}
  \end{align}
Equation~\eqref{qccoupgreen} is thus  reduced to the more compact form:
 \begin{align}
&\quad  \eta_L \left ( 1+g_L^{(1)}  g_L^{(2)}   \right )  \cos \left (\delta_L^{(1)}   - \delta_L^{(2)}    \right )  \nonumber \\
& + \left ( 1-g_L^{(1)}  g_L^{(2)}   \right )   \cos \left (\delta_L^{(1)}   + \delta_L^{(2)}    \right )   \nonumber \\
& -  \eta_L  \left (g_L^{(1)}  - g_L^{(2)}   \right ) \sin \left (\delta_L^{(1)}   - \delta_L^{(2)}    \right )   \nonumber \\
& - \left ( g_L^{(1)} + g_L^{(2)}   \right )  \sin \left (\delta_L^{(1)}   + \delta_L^{(2)}    \right ) =0, \label{coupledQC}
\end{align}
where
\begin{equation}
g_L^{(\alpha)} =   (-1)^{L+1} \left (  \frac{ 4\mu_\alpha \omega}{q_\alpha^2} \right )^{L+ \frac{1}{2}}         \frac{ \Gamma (\frac{L}{2} + \frac{3}{4} - \frac{ \epsilon^{( \omega)}_\alpha }{2 \omega}) }{\Gamma ( \frac{1}{4}-\frac{L}{2} - \frac{\epsilon^{( \omega)}_\alpha}{2\omega} )}. \label{gLfunc}
\end{equation}
The  coupled-channel QC in a harmonic trap expressed by
Eq.(\ref{coupledQC}) resembles that in 
a periodic cube [see Eq.(25) in \cite{Guo:2012hv}]. In the harmonic trap, $ g_L^{(\alpha)} $ functions play the same role of L\"uscher  zeta functions in the periodic cube.

Given one energy eigenvalue $E$, we can not determine all of $\delta_L^{(1, 2)} $ and $\eta_L $ simultaneously. A  possible way out  is to  gather   several  pairs of  $( E, \omega)$    to build more constraints on $\delta_L^{(1, 2)} $ and $\eta_L $, provided that energy eigenvalues $E$'s are somewhat close to each other and the ratio $E/\omega$ is distinct from one another. Though this approach is  model- independent, the requirement may only be met by small portion of the calculated levels. 
Another useful approach widely used in the lattice QCD community is to  model $\delta_L^{(1, 2)} $ and $\eta_L $ with a few free parameters,
which can be fitted to more energy levels
by using QC as defined in Eq.(\ref{coupledQC}), 
(see, e.g., discussions in Refs.~\cite{Guo:2012hv,Guo:2013vsa}).

\section{Discussions and Summary}\label{summary}

The harmonic trap may be replaced with artificial fields of different configuration or boundary condition. 
For instance, hard-sphere boundary condition is used in some Monte Carlo simulations~\cite{Bovermann:2019jbt,Elhatisari:2016hby,Rokash:2015hra}.
We can repeat all the discussions in previous sections by substituting the following external potential in place of the HO potential:
\begin{equation}
V_{h.s.} (r)= \begin{cases} 0, & r<b  \, , \\ \infty, & r>b\, ,  \end{cases}
\end{equation}
where $b$ is the radius of the sphere.  The Green's function must satisfy the boundary condition:
\begin{align}
&   G_{L}^{(h.s.)} (r, r'; q)  \nonumber \\
& = - 2\mu  q j_L( q r_<) j_L (q r_>) \left [ \frac{n_L (q b)}{j_L (q b)}- \frac{n_L (q r_>)}{j_L (q r_>)} \right ] . 
\end{align}
For instance, in the case of the single-channel elastic scattering,
Eqs. (\ref{eqn:invTsingle}) and (\ref{eqn:homoVsingle}) are replaced with
\begin{align}
&    \frac{2^{2L+2} \Gamma^2(L+\frac{3}{2})}{\pi} \frac{G^{(h.s.)}_L (r,r'; q )- G^{(\infty)}_L ( r,r' ; q  ) }{(r r')^L} |_{r,r' \rightarrow 0}     \nonumber \\
 &  = - 2\mu q^{2L+1} [  \cot \delta_L (q  ) - i ].
\end{align}
Using the asymptotic form of $G^{(\infty)}_L$ (\ref{asymp2}) and
\begin{align}
 &   \frac{1}{2\mu}   \frac{2^{2L+2}\Gamma^2(L + \frac{3}{2}) }{\pi }     \frac{G^{(h.s.)}_{ L} (r,r'; q  )  }{(r r')^L} |_{r,r' \rightarrow 0}  \nonumber \\
   &  =  - q^{2 L+1}         \frac{n_L (q  b)}{j_L (q b)}   -  \frac{2^{2L+1}  \Gamma(L+ \frac{1}{2}) \Gamma(L + \frac{3}{2})}{\pi  } \frac{1}{r^{2L+1}}    |_{r\rightarrow 0},
\end{align}
the quantization condition of the single-channel case for the spherical hard wall is thus given by a form analogous to the BERW formula (or  L\"uscher's for that matter):
 \begin{align}
  \cot \delta_L (q  )   =  \frac{n_L (q  b)}{j_L (q  b)}      . \label{qchardwall}
\end{align}

The hard-sphere geometry is encoded in $  n_L (q  b)/j_L (q  b)  $, which plays the same role as $ g_L (q) $ to the harmonic trap.
Similarly, the coupled-channel QC can be worked out for the spherical hard wall by
replacing $g_L^{(\alpha)} $ in Eq.(\ref{gLfunc}) as follows
\begin{equation}
g_L^{(\alpha)}  \rightarrow   \frac{n_L (q_\alpha b)}{j_L (q_\alpha b)}   \, .
\end{equation}

In summary, we have derived the quantization condition for coupled-channel two-particle systems trapped by the harmonic potential. The formula by itself relates the coupled-channel reaction amplitudes in infinite volume, parametrized by two phase shifts and one inelasticity, to energy eigenvalues of the corresponding trapped system. Although only the quantization condition for the case of distinguishable spinless particles is shown, with the technical detail spelled out in the paper, the extension to include slight complications due to spin, identical particles and so on should be straightforward. Coulomb repulsion is important, however, if reactions take place near the threshold of charged particles. Quantization conditions may be derived for that circumstance in the harmonic trap, following the methodology developed in Ref.~\cite{Guo:2021lhz}.

\acknowledgments

B.L. thanks Sebastian K\"onig and Baishan Hu for useful discussions at early stage of this work. P.G. acknowledges support from the Department of Physics and Engineering, California State University, Bakersfield, CA. The work (BL) was supported in part by the National Natural Science Foundation of China (NSFC) under Grants No.11775148 and No.11735003, and (PG) was also supported in part by the National Science Foundation (US) under Grant No. NSF PHY-1748958.

\bibliography{ALL-REF.bib,nuclph.bib}

\end{document}